\documentclass[aps,pra,groupedaddress,reprint]{revtex4-1}
\usepackage{graphicx}
\usepackage{dcolumn}
\usepackage{bm}
\usepackage{amsmath}
\allowdisplaybreaks

\begin{document}

\title{High Visibility Two-photon Interference with Classical Light}

\author{Peilong Hong}
\affiliation{The MOE Key Laboratory of Weak Light Nonlinear Photonics and School of Physics, Nankai University, Tianjin 300457, China}

\author{Lei Xu}
\affiliation{The MOE Key Laboratory of Weak Light Nonlinear Photonics and School of Physics, Nankai University, Tianjin 300457, China}

\author{Zhaohui Zhai}
\affiliation{The MOE Key Laboratory of Weak Light Nonlinear Photonics and School of Physics, Nankai University, Tianjin 300457, China}

\author{Guoquan Zhang}
\email{zhanggq@nankai.edu.cn}
\affiliation{The MOE Key Laboratory of Weak Light Nonlinear Photonics and School of Physics, Nankai University, Tianjin 300457, China}

\date{\today}

\begin{abstract}
Two-photon interference with independent classical sources, in which superposition of two indistinguishable two-photon paths plays a key role, is of limited visibility of interference fringes with a maximum value of 50\%. By using a random-phase grating to modulate the wavefront of a coherent light, we introduce superposition of multiple indistinguishable two-photon paths, which enhances the two-photon interference effect with a signature of visibility exceeding 50\%. The result shows the importance of phase control in the control of high-order coherence of classical light.

\end{abstract}

\maketitle

\section{Introduction}

Interference is an essentially important topic in optical physics, resulting in many interesting phenomena and important applications. The key physics lying behind interference is the superposition principle. After the birth of quantum physics, it was realized that superposition of multiple single-photon paths plays a key role in the traditional optical interference phenomenon, which is usually known as Dirac's famous statement:  \emph{``Each photon interferes only with itself. Interference between different photons never occurs''}~\cite{dirac1935principles}.

\begin{figure}[htb]
\centering
\includegraphics[width=0.4\textwidth]{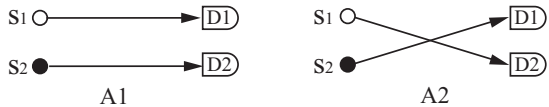}
\caption{Two indistinguishable two-photon paths for a pair of independent photons S1 and S2 to trigger a coincidence count. D1 and D2 are two single-photon detectors.\label{path1}}
\end{figure}

In 1956, Hanbury Brown and Twiss introduced the second-order correlation measurement, and reported a new type of interference effect between independent photons, i.e., the bunching effect of thermal light~\cite{brown1956correlation}. Soon after, it was realized that superposition of two indistinguishable two-photon paths plays a key role in this type of interference effect~\cite{fano1961quantum}. As shown in Fig.~\ref{path1}, for every pair of independent photons, there are two indistinguishable paths for the pair of photons to trigger a coincidence count, and the amplitudes of these two paths are always of the same random phase $\phi_{s1}+\phi_{s2}$, where $\phi_{si} (i=1,2)$ is the random phase of the point source emitting photons $si$. Therefore, the interference term between the two amplitudes will survive in the ensemble average, leading to the constructive or destructive two-photon interference.

Later, Mandel gave a detailed theoretical analysis of two-photon interference between two independent sources with random phases, and predicted that the visibility of two-photon interference fringes has a maximum value of 50\% for classical light~\cite{mandel1983photon}. In general, for the case of two-photon interference with independent light sources,  the coincidence counts consist of two parts: (i) The self-correlation part with the pair of photons from the same source which contributes a constant to the correlation function. (ii) The cross-correlation part with the pair of photons from different sources, which is dominated by the superposition of the two indistinguishable paths as depicted in Fig.~\ref{path1}, resulting in the two-photon interference fringes. It is the existence of part (i) that will low down the visibility of two-photon interference fringes. For a quantum source such as the single-photon state, the self-correlation contribution could be eliminated, and therefore giving rise to a 100\%-visibility two-photon interference~\cite{mandel1983photon}. However, for a classical light, the self-correlation always contributes, which makes the visibility of two-photon interference fringes not exceeding 50\%. This property of two-photon interference with independent random-phase sources was further discussed and demonstrated by Paul~\cite{paul1986interference}, Ou~\cite{ou1988quantum} and Klyshko~\cite{klyshko1994quantum}. Nevertheless, the visibility of multi-photon interference fringes of classical lights based on third- and higher-order coherence could be higher than 50\%~\cite{agafonov2008high,cao2008enhancing,chen2010high,zhou2010third}.

In this paper, without the introduction of third- and higher-order correlation measurement, we explore another way to achieve high visibility two-photon interference with classical light. Instead of reducing or even removing the self-correlation contribution as in the case of quantum sources, we increase the cross-correlation contribution by introducing the superposition of multiple indistinguishable two-photon paths (path number $>$2) to enhance the two-photon interference effect of a classical light. This could be practically realized by introducing a random-phase grating to modulate the wavefront of a coherent light.

\section{Random-phase Grating}

\begin{figure}[!htb]
\centering
\includegraphics[width=0.4\textwidth]{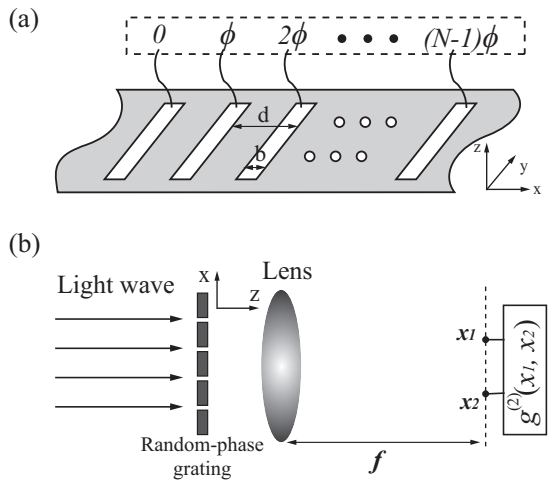}
\caption{(a) Schematic diagram for the designed random-phase grating with $N$ slits. The inset shows the random phases encoded on the light waves transmitting through the respective slits of the grating, in which the elementary phase $\phi$ changes with time randomly. (b) Schematic diagram for detecting the
two-photon interference of the light wave transmitting through the $N$-slit random-phase grating in the Fraunhofer zone, where $f$ is the focal length of the lens. \label{grating}}
\end{figure}

The random-phase grating is shown schematically in Fig.~\ref{grating}(a). It is a transmission $N$-slit mask with specially designed random-phase structure shown in the inset of Fig.~\ref{grating}(a). The transmission slit width is $b$ and the distance between neighboring slits is $d$. A random phase $(n-1)\phi$ will be encoded on the light wave transmitting through the $nth$ slit, where $n$ is an integer and the phase $\phi$ changes randomly with time. Such a random-phase grating can be realized through a spatial light modulator (SLM) in practice, as we will demonstrate in the following. It is evident that there will be no stationary first-order interference for a coherent light transmitting through the random-phase grating since the phase difference between every two slits changes randomly with time. However, the case will be totally different when one considers the two-photon interference. Here we will consider the case that a collimated coherent light transmitting through the random-phase grating, as shown in Fig.~\ref{grating}(b).

Now, let's consider the two-photon paths in the scheme that contribute to the
two-photon interference. For each pair of slits, the $mth$ and $nth$ slits, of the random-phase grating, there are always two indistinguishable two-photon paths as shown in Fig.~\ref{path1}: (1) one photon transmitting through the $mth$ slit goes to the detector D1, while the other transmitting through the $nth$ slit goes to the detector D2; and (2) one photon transmitting through the $mth$ slit goes to the detector D2, while the other transmitting through the $nth$ slit goes to the detector D1. Note that the same random phase $(m+n-2)\phi$ is encoded on the amplitudes of the two different but indistinguishable two-photon paths. For a $N$-slit random-phase grating as shown in Fig.~\ref{grating}, one may find that there are many such twin two-photon paths originated from different pairs of slits ($m$, $n$), and the amplitudes of those twin paths with equal ($m+n$) will contain the same random phase $(m+n-2)\phi$. These twin two-photon paths are indistinguishable in principle. In this way, we introduce multiple different but indistinguishable two-photon paths through the $N$-slit random-phase grating, and the superposition of these multiple two-photon amplitudes would enhance the two-photon interference, leading to high-visibility two-photon interference for classical light.

\section{Theoretical Results}

To clearly illustrate the two-photon interference effect of the light field transmitting through the random-phase grating, we will calculate the second-order correlation function of the transmitting field in the Fraunhofer zone, i.e., in the focal plane of a lens put behind the random-phase grating, as shown in Fig.~\ref{grating}(b). For simplicity, we assume that the coherent light incident on the random-phase grating is a plane wave and a single-mode one, in one dimensional case, the field operator on the detection plane is expressed as~\cite{glauber1963quantum,brooker2003modern}
\begin{equation}
\hat{E}^{(+)}(x_j)\propto \int_{-\frac{b}{2}}^{\frac{b}{2}} e^{-i\beta_j x_s} \mathrm{d} x_s \sum_{n=0}^{N-1} e^{-in(\beta_j d-\phi)} \hat{a}\, .\label{field}
\end{equation}
Here $\hat{a}$ is the annihilation operator, $\beta_j=k \sin\theta_j$ and $\tan\theta_j=x_j/f$ $(j=1,2)$ with $f$, $k$ and $\theta_j$ being the focal length of the lens, the wave vector and the diffraction angle of the light wave, respectively.

\emph{First-order Spatial Correlation Function} --- The first-order spatial correlation function in the detection plane can be expressed as~\cite{glauber1963quantum}
\begin{equation}
G^{(1)}(x_j,x_j)=\langle E^*(x_j) E(x_j)\rangle \, ,\label{g1}
\end{equation}
where $E^*(x_j)$ is the eigenvalue of the field operator $\hat{E}^{(+)}(x_j)$ on the state of source (coherent state), and $\langle \cdots \rangle$ represents the ensemble average. By substituting Eq.~(\ref{field}) into Eq.~(\ref{g1}), one gets
\begin{equation}
G^{(1)}(x_j,x_j)\propto N\left(\frac{\sin(\beta_j b/2)}{\beta_j b/2}\right)^2 \, .\label{g1r}
\end{equation}
Here the term $\sin^2(\beta_j b/2)/(\beta_j b/2)^2$ represents the diffraction from a single slit. It is evident that the intensity distribution in the focal plane is a sum of the diffraction intensities from $N$ different slits. There is no stationary first-order interference among these slits of the random-phase grating.

\emph{Second-order Spatial Correlation Function} --- The second-order spatial correlation function at the detection plane can be expressed as~\cite{glauber1963quantum}
\begin{equation}
 G^{(2)}(x_1,x_2)=\langle E^*(x_1) E^*(x_2) E(x_1) E(x_2) \rangle\, .\label{g2}
\end{equation}
By substituting Eq.~(\ref{field}) into Eq.~(\ref{g2}), the second-order spatial correlation function can be deduced as
\begin{widetext}
 \begin{equation}\label{path}
 \begin{split}
G^{(2)}(x_1,x_2) \propto&\sum_{l=0}^{2N-2} \Big\arrowvert \prod_{j=1}^{2} \int_{-\frac{b}{2}}^{\frac{b}{2}} e^{-i\beta_j x_s}\mathrm{d}x_s \times \sum_{(m,n);m+n=l}(e^{-in(\beta_1 d-\phi)}
    \cdot e^{-im(\beta_2 d-\phi)}\\
    &+(1-\delta(m-n)) e^{-in(\beta_2 d-\phi)} e^{-im(\beta_1 d-\phi)})\Big\arrowvert ^2\\
  \propto & \prod_{j=1}^{2} \left(\frac{\sin(\beta_j b/2)}{\beta_j b/2}\right)^2 \times \sum_{l=0}^{2N-2} \Big\arrowvert \sum_{(m,n);m+n=l} (e^{-in\beta_1 d} e^{-im\beta_2 d}
  \quad+(1-\delta(m-n))\cdot e^{-in\beta_2 d} e^{-im\beta_1 d})\Big\arrowvert ^2,
 \end{split}
 \end{equation}
 \end{widetext}
Note that the terms $e^{-in\beta_1 d} e^{-im\beta_2 d}$ and $e^{-in\beta_2 d} e^{-im\beta_1 d}$ correspond to the twin two-photon paths transmitting through the $mth$ and $nth$ slits of the random-phase grating, respectively. Here we introduce the delta function $\delta(m-n)$ to show that there is only one path when the two photons transmit through the same slit to trigger a coincidence count. Following the superposition principle, in Eq.~(\ref{path}), the amplitudes of all different but indistinguishable two-photon paths with the same random phase $l\phi$ ($l=0,\, 1,\cdots,\, 2N-2$) are superposed to calculate their contributions to the coincidence probability, and then the coincidence probability contributions from those with different random phases $l\phi$ are added to get the total coincidence probability.

\begin{figure}[!htb]
\centering
\includegraphics[width=0.4\textwidth]{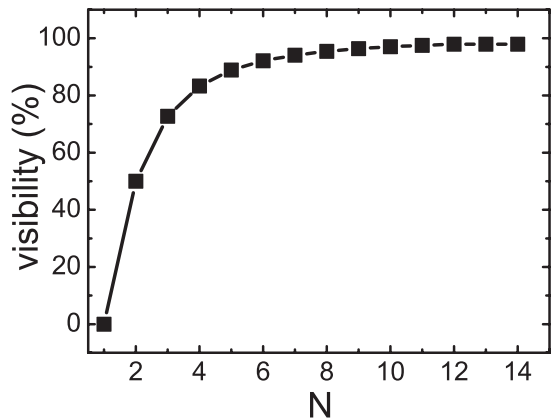}
\caption{The dependence of the two-photon interference visibility $V$ on the slit number $N$ of the random-phase grating.\label{vis}}
\end{figure}

Thus, the normalized second-order spatial correlation function can be calculated as
 \begin{equation}\label{g2r}
 \begin{split}
 g^{(2)}(x_1,x_2)=&\frac {G^{(2)}(x_1,x_2)}{G^{(1)}(x_1,x_1)G^{(1)}(x_2,x_2)} \\
   =&\frac{1}{N^2}\sum_{l'=-N}^{N-1}\frac{\sin^2((l'+1)(\beta_1-\beta_2)d/2)}{\sin^2((\beta_1-\beta_2)d/2)}\, .
 \end{split}
 \end{equation}
One notes that the terms $\sin^2((l'+1)(\beta_1-\beta_2)d/2)/\sin^2((\beta_1-\beta_2)d/2)$ are of the similar formula as the multiple-slit single-photon interference function~\cite{brooker2003modern}, and therefore can be called as multiple-slit two-photon interference function. It is seen that $g^{(2)}(x_1,x_2)$ in Eq.~(\ref{g2r}) is a sum of $(2N-1)$ multiple-slit two-photon interference functions $\sin^2((l'+1)(\beta_1-\beta_2)d/2)/\sin^2((\beta_1-\beta_2)d/2)$ introduced by the random-phase grating, each one is associated with a group of different but indistinguishable two-photon paths which are characterized by the same random phase $l\phi$ ($l=0,\, 1,\cdots,\, 2N-2$) in Eq.~(\ref{path}). These multiple-slit two-photon interference functions are peaked at the same condition $(\beta_1-\beta_2)d=\pm 2n\pi$ ($n=0$, $1$, $2$, $\cdots$) due to the constructive interference effect, i.e., when the phase difference among different but indistinguishable two-photon paths are an integer multiple of $2\pi$. The constructive interference peak for each multiple-slit two-photon interference function is $(l'+1)^2$, and therefore, one can get the interference peak of $g^{(2)}(x_1,x_2)$ to be $(2N^2+1)/(3N)$, according to Eq.~(\ref{g2r}). It can also be easily seen from Eq.~(\ref{g2r}) that $g^{(2)}(x_1,x_2)$ is a periodical function of the position difference ($x_1-x_2$) with a period of $\Lambda=\lambda f/d$ in the paraxial approximation, which is exactly the same as that of the multiple-slit single-photon interference pattern of a normal grating with respect to the position $x$ on the detection plane~\cite{brooker2003modern}.
On the other hand, the minimum of $g^{(2)}(x_1,x_2)$ is achieved at the condition $(\beta_1-\beta_2)d=\pm (2n+1)\pi$ ($n=0$, $1$, $2$, $\cdots$) due to the destructive interference effect among multiple two-photon paths. However, the minimum of $g^{(2)}(x_1,x_2)$ is not zero but calculated to be $1/N$ due to the existence of the cases when the two photons transmit through the same slit of the grating. Therefore, the visibility of the two-photon interference, defined as $V=(g^{(2)}_{max}-g^{(2)}_{min})/(g^{(2)}_{max}+g^{(2)}_{min})$, is $(N^2-1)/(N^2+2)$, which grows quickly with the increase of slit number $N$ and exceeds 50\% when $N>2$, as shown in Fig.~\ref{vis}.

\section{Experimental Demonstration and Discussions}

\begin{figure}[!htb]
\centering
\includegraphics[width=0.45\textwidth]{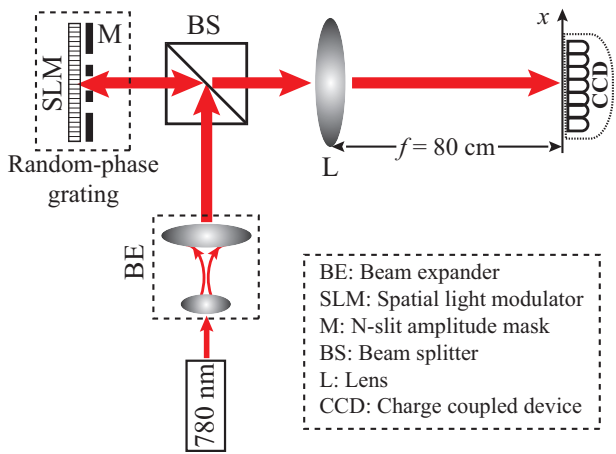}
\caption{Schematic diagram of the experimental setup. A single mode 780-nm laser was introduced as the light source. A $N$-slit amplitude mask ($N=2$, 3, 4 and 5, $b = 72$ $\mu$m and $d=400$ $\mu$m) and a SLM were used to construct the random-phase grating shown in Fig.~\ref{grating}. A CCD camera was located at the focal plane of the lens L to measure the intensity distribution and the second-order spatial correlation function. \label{setup}}
\end{figure}

\emph{Experimental Setup} --- Figure~\ref{setup} shows the experimental setup that we used to measure the two-photon interference effect of the light field scattering from the random-phase grating. In our experiments, a single mode laser with a wavelength of 780 nm was introduced as the light source, which was expanded and collimated through a beam expander to obtain a plane wave. The expanded and collimated light beam was then reflected by a beam splitter BS and incident normally onto a random-phase grating. Here the random-phase grating was composed of a $N$-slit amplitude mask ($b=72$ $\mu$m and $d=400$ $\mu$m) and a reflection-type phase-only SLM (HEO 1080P from HOLOEYE Photonics AG, Germany) which was put as close as possible to the mask. The light first transmitted through the $N$-slit amplitude mask, and then was reflected back from the SLM and finally re-transmitted through the $N$-slit amplitude mask again. The SLM provided the desired random-phase structure on every individual slit of the $N$-slit amplitude mask as shown
in the inset of Fig.~\ref{grating}(a).
At last, the light waves scattered from the random-phase grating were collected by a lens L with a focal length $f=80$ cm. Both the intensity and the second-order spatial correlation measurements were performed on the focal plane of the lens L by using a charge coupled device (CCD) camera.

\begin{widetext}

\begin{figure}[!htb]
\centering
\includegraphics[width=0.84\textwidth]{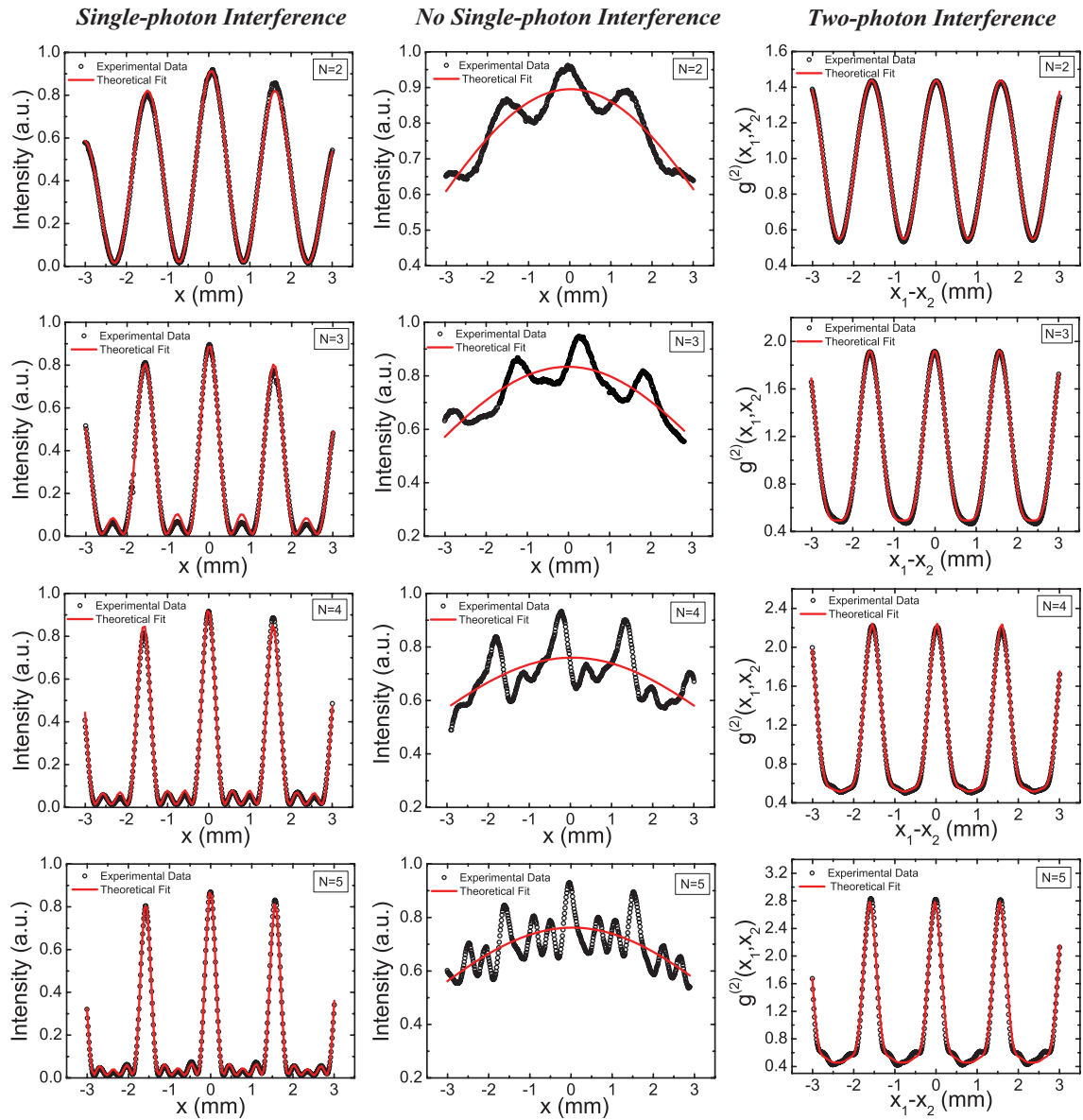}
\caption{The single-photon interference and the two-photon interference of light on the detection plane at different conditions. The empty circles are the experimental data, while the red solid curves are the theoretical fits. The first column shows the stationary single-photon interference fringes with the normal $N$-slit gratings. The second-column shows the averaged intensity distributions over 10000 frames of instantaneous  single-photon interference patterns with the $N$-slit random-phase gratings. And the third column represents the two-photon interference fringes with the $N$-slit random-phase gratings. Here we used a CCD camera to measure the instantaneous intensity distributions on the detection plane, as shown in Fig.~\ref{setup}. In all cases, the slit number $N$ of the grating was set to be 2, 3, 4 and 5, respectively.\label{interference}}
\end{figure}

\end{widetext}

Figure~\ref{interference} shows the measured single-photon and two-photon interference patterns on the detection plane (i.e., the focal plane of the lens L) at different conditions, in which the empty circles are the experimental results while the red curves are the theoretical fits, respectively.

\emph{Results for Traditional Grating} --- When there is no electric signal loaded on the SLM, our experimental configuration is essentially the same as a typical setup to measure the single-photon interference of a traditional $N$-slit grating.
In the experiment, we measured the single-photon interference patterns of the $N$-slit gratings ($N=2$, 3, 4 and 5, respectively). The results are shown in the first column of Fig.~\ref{interference}.
As expected, stationary single-photon interference fringes described by the multiple-slit single-photon interference function $\sin^2(N \beta d/2)/\sin^2(\beta d/2)$~\cite{brooker2003modern} were observed. The period between the neighboring principal intensity peaks was measured to be 1.57 mm on the detection plane, and $(N-2)$ sub-peaks appear between the two neighboring principal peaks of the stationary single-photon interference fringes. Note that the normalized second-order spatial correlation function $g^{(2)}(x_1,x_2)$ in this case was confirmed to be a unity (not shown in Fig.~\ref{interference}).

\emph{Results for Random-phase Grating} --- When the SLM was loaded with the random phases, the random-phase grating was constructed. In this case, there should be no stationary single-photon interference fringes since the phase difference between every two slits changes randomly.  The second column of Fig.~\ref{interference} shows the experimental results.
One can find that the single-photon interference fringes were erased, leaving
an intensity distribution enveloped by the single-slit diffraction profile as described by Eq.~(\ref{g1r}). One notes that there are still some residual intensity fluctuations which deviate from the intensity envelop described by Eq.~(\ref{g1r}).
This is mainly due to the unavoidable phase flicker effect of the SLM, which leads to imprecise control on the random phases on the slits of the grating.

Interestingly, although the single-photon interference fringes disappear with the random-phase grating, two-photon interference fringes appear. As shown in the third column of Fig.~\ref{interference}, the second-order spatial correlation function $g^{(2)}(x_1,x_2)$ exhibits itself in the form of high quality interference fringes, which is in good agreement with the theoretical prediction by Eq.~(\ref{g2r}). Here the second-order correlation function is calculated through a formula $g^{(2)} (x_1, x_2)=\langle I(x_1)I(x_2) \rangle/(\langle I(x_1)\rangle\langle I(x_2)\rangle)$ by using 10000 frames of the instantaneous intensity distributions measured by the CCD camera~\cite{chen2010high,CCD}.
It is seen from the experimental results shown in the third column of Fig.~\ref{interference} that, the two-photon interference fringes are peaked at the position differences $x_1-x_2=\pm 2n\pi f/(kd)$ but minimized at the position differences $x_1-x_2=\pm (2n+1)\pi f/(kd)$ ($n=0$, $1$, $2$, $\cdots$), respectively. The period of the two-photon interference fringes $\Lambda$ was measured to be 1.57 mm, in good agreement with the prediction of Eq.~(\ref{g2r}). On the other hand, sub-peaks typical for the single-photon interference fringes shown in the first column of Fig.~\ref{interference} were not observed in the two-photon interference fringes in the third column of Fig.~\ref{interference}. This is due to the fact that $g^{(2)}(x_1,x_2)$ is a sum of $(2N-1)$ different multiple-slit two-photon interference functions (see Eq.~(\ref{g2r})), and  these different multiple-slit two-photon interference functions are always in phase at their principal peaks but out of phase at the sub-peaks. Moreover, the visibility of the two-photon interference fringes was measured to be $44.9\%,\, 59.1\%,\, 62.3\%$ and $71.9\%$ for the $N$-slit random-phase gratings with $N=2,\,3,\, 4$ and $5$, respectively.
As predicted by Eq.~(\ref{g2r}), the visibility of the two-photon interference fringes increases with the increase of the slit number $N$ of the random-phase gratings and surpasses 50\% when $N>2$.

\emph{Further Discussions} --- One may note that the random phases encoded on the slits of the random-phase grating are not fully independent but indeed correlated with respect to each other. Therefore, our case is different from the case discussed by Mandel~\cite{mandel1983photon}, where classical lights with fully independent random phases are considered. For classical lights with fully independent random phases, the visibility of two-photon interference fringes cannot exceed 50\%. While by appropriately controlling the random phase structure encoded on a coherence light field, one could realize two-photon interference with its visibility exceeding 50\%. It is known that controlling optical phase plays a key role in the single-photon interference effect, our results show that it may also play an important role in controlling the high-order coherence of light.

\section{Summary}

In summary, we have designed a kind of two-photon grating with a special random-phase structure, through which the single-photon interference is smeared out but the two-photon interference appears. With such a random-phase grating, superposition of multiple indistinguishable two-photon paths is introduced, which leads to a high visibility two-photon interference fringes of classical light. Theoretically, the visibility of the two-photon interference fringes for a coherent light transmitting through a $N$-slit random-phase grating reaches $(N^2-1)/(N^2+2)$. Experimentally, the visibility of the two-photon interference fringes with a $N$-slit random-phase grating ($N =2,\,3,\,4$ and 5) was measured to be $44.9\%,\, 59.1\%,\, 62.3\%$ and $71.9\%$, respectively.  The results show the possibility to control the high-order coherence of light through optical phase.

\section*{Acknowledgements}

This work was supported by the 973 program (2013CB328702), the CNKBRSF (2011CB922003), the NSFC (11174153, 90922030 and 10904077), the 111 project (B07013), and the Fundamental Research Funds for the Central Universities.


\begin{thebibliography}{99}
\newcommand{\enquote}[1]{``#1''}

\bibitem{dirac1935principles}
P.~Dirac, \emph{The principles of quantum mechanics. 2nd edition} (Oxford:
  Clarendon Press, 1935).

\bibitem{brown1956correlation}
R.~Brown and R.~Twiss, Nature \textbf{177}, 27 (1956); \textbf{178}, 1046 (1956).

\bibitem{fano1961quantum}
U.~Fano, Am. J. Phys. \textbf{29}, 539 (1961).

\bibitem{mandel1983photon}
L.~Mandel, Phys. Rev. A \textbf{28}, 929 (1983).

\bibitem{paul1986interference}
H.~Paul, Rev. Mod. Phys. \textbf{58}, 209 (1986).

\bibitem{ou1988quantum}
Z.~Ou, Phys. Rev. A \textbf{37}, 1607 (1988).

\bibitem{klyshko1994quantum}
D.~Klyshko, Phys. Usp. \textbf{37}, 1097 (1994).

\bibitem{agafonov2008high}
I.~Agafonov, M.~Chekhova, T.~Iskhakov, and A.~Penin, Phys. Rev. A \textbf{77},
  053801 (2008).

\bibitem{cao2008enhancing}
D.~Cao, J.~Xiong, S.~Zhang, L.~Lin, L.~Gao, and K.~Wang, Appl. Phys. Lett.
  \textbf{92}, 201102 (2008).

\bibitem{chen2010high}
X.~Chen, I.~Agafonov, K.~Luo, Q.~Liu, R.~Xian, M.~Chekhova, and L.~Wu, Opt.
  Lett. \textbf{35}, 1166 (2010).

\bibitem{zhou2010third}
Y.~Zhou, J.~Simon, J.~Liu, and Y.~Shih, Phys. Rev. A \textbf{81}, 043831
  (2010).

\bibitem{glauber1963quantum}
R.~Glauber, Phys. Rev. \textbf{130}, 2529 (1963); \textbf{131}, 2766 (1963).

\bibitem{brooker2003modern}
G.~Brooker, \emph{Modern classical optics} (Oxford University Press Oxford, UK,
  2003).

\bibitem{CCD}
Y. Bromberg, Y. Lahini, E. Small, and Y. Silberberg, Nature Photonics \textbf{4}, 721 (2010).

\end{thebibliography}
\end{document}